\documentclass[preprint]{aastex631}
\usepackage{booktabs}
\usepackage{soul}

\accepted{October 22, 2024}

\submitjournal{PASP}

\begin{document}

\title{CAMO-S: A meteor-tracking spectrograph at the Canadian Automated Meteor Observatory}

\correspondingauthor{Michael Mazur}
\email{mmazur5@uwo.ca}

\author[0000-0002-7662-9046]{Michael J. Mazur}
\affiliation{Department of Earth Sciences \\
Western University \\
1151 Richmond St. \\
London, Ontario \\
Canada N6A 3K7}

\author{Margaret Campbell-Brown}
\affiliation{Department of Physics and Astronomy \\
Western University \\
1151 Richmond St. \\
London, Ontario \\
Canada, N6A 3K7}
\affiliation{Institute for Earth and Space Exploration \\
Western University \\
1151 Richmond St. \\
London, Ontario \\
N6A 5B8, Canada}

\author{Peter G. Brown}
\affiliation{Department of Physics and Astronomy \\
Western University \\
1151 Richmond St. \\
London, Ontario \\
Canada, N6A 3K7}
\affiliation{Institute for Earth and Space Exploration \\
Western University \\
1151 Richmond St. \\
London, Ontario \\
N6A 5B8, Canada}

\author{Denis Vida}
\affiliation{Department of Physics and Astronomy \\
Western University \\
1151 Richmond St. \\
London, Ontario \\
Canada, N6A 3K7}
\affiliation{Institute for Earth and Space Exploration \\
Western University \\
1151 Richmond St. \\
London, Ontario \\
Canada, N6A 5B8}

\author{Pete Gural}
\affiliation{Gural Software and Analysis LLC \\
351 Samantha Dr. \\
Sterling, Virginia \\
USA, 20164}

\author{Zhangqing Yang}
\affiliation{Department of Physics and Astronomy \\
Western University \\
1151 Richmond St. \\
London, Ontario \\
Canada, N6A 3K7}

\begin{abstract}

The Canadian Automated Meteor Observatory (CAMO) mirror tracking system has been in operation since 2009 and has, to date, produced more than 20,000 two-station meteor observations at meter-level spatial and 10 ms temporal resolution. In 2020, a spectral tracking camera was added in parallel at one of the CAMO stations. To date, it has recorded the spectra of hundreds of faint meteors. Engineering testing from 2020-2023 resulted in the selection of a 150 lpmm grating and an EMCCD camera to achieve a spectral resolution of about 1 nm/pixel in the final configuration. The CAMO spectral system can resolve spectra from individual meteoroid fragments, record spectra for meteors of +2 peak magnitude to as faint as +4 in parts of the lightcurve and produce relative abundance estimates for Mg, Fe and Na. Our preliminary results also show identification of the H and K lines of CA(II). Meteors with strong iron lines were found to have unusual fragmentation behaviour, involving gross fragmentation rather than continuously shedding small particles. The spectra of individual fragments can be resolved in some cases, showing that these Fe-rich objects do not differ in composition among fragments. Our calibration procedure and hardware configuration are discussed together with preliminary results. 

\end{abstract}


\section{Introduction}

As a meteoroid ablates in the atmosphere, the dense plasma surrounding and trailing behind the meteoroid produces a distinctive spectrum which predominantly originates from ablated meteoroid atoms. This spectrum can be related to  the original meteoroid's bulk chemistry \citep{Ceplecha1998} through application of appropriate models \citep{Borovicka1993}. When this spectrum is juxtaposed with a meteoroid's pre-atmospheric orbit, it offers a robust method to link meteoroid populations with their respective compositions \citep{Vojacek2015}.

Spectral meteor research progressed significantly in the mid-1990s, with high-resolution fireball spectra \citep{Borovicka1993} and exploration of the nexus between meteor spectra, meteorite classes, and asteroid families \citep{Borovicka1994a}. A major advance was work by \citet{Borovicka1994b} which revealed that meteor spectra can be modelled with two distinct thermal components, each of which have relative intensities significantly influenced by meteoroid velocity . Presently, only a few digital surveys of meteor spectra exist, with notable examples being from  \citet{Borovicka2005}, \citet{Matlovic2019} and \citet{Vojacek2015}. A recent noteworthy study by \cite{Devillepoix2022} focused on meteorite-dropping fireballs observed in Australia's Nullarbor region. Their findings shed light on the spectra of chondritic bodies with pronounced Fe, Mg, and Cr lines. 

The purpose of this work is to describe the construction, testing and initial results of a sensitive, high-spatial resolution tracking video system designed to study the spectra and fragmentation of meteors of moderate brightness. To date, most published meteor spectral work has been done on meteors brighter than magnitude +2 \citep{Borovicka2019a}; while the majority of described meteor spectra are from significantly brighter events. Clearly there is need to augment spectral surveys to fainter magnitudes. This will help extend the chemical and physical correlations among small body populations which have already been examined at larger (cm) sizes \citep[eg.][]{Vojacek2018, Matlovic2019}.

 In addition to the current lack of high-quality spectra of faint meteors, another key element that has been missing from prior surveys is the ability to distinguish between wake and fragment spectra. To do this, a high spatial and temporal resolution system is required - one that can resolve individual fragments of meteors as they disintegrate in the atmosphere. With these goals in mind, leveraging the existing imaging and tracking capabilities of the Canadian Automated Meteor Observatory \citep{Vida2021} suggested it would be an ideal platform for studying not only the optical properties of faint fragmenting meteors but also their spectra.

\section{Equipment and the Canadian Automated Meteor Observatory}

The goals of the new spectral system are to track faint meteors with a spectral camera and provide relative (and ideally absolute) abundances; to resolve the spectra of individual fragments or of the head and wake of meteors; and to identify the presence or absence of refractory inclusions among meteor fragments.  While the first goal is relatively easily achieved, the second and third are more challenging. Being able to identify spectra from individual fragments will make it possible to compare their spectral signatures and identify gross compositional differences between them.

With these goals in mind, the new high-resolution spectral system was chosen to be an augmentation of the existing instrument suite at one of the two stations comprising the Canadian Automated Meteor Observatory (CAMO). CAMO consists of two facilities - one at the Elginfield Observatory (81.31566 W, 43.19279 N) and one co-located with the Canadian Meteor Orbit Radar (CMOR) \citep{Brown2008} near Tavistock, Ontario, Canada (80.77209 W, 43.26420 N). 

The CAMO mirror tracking system consists of a wide-field fixed intensified CCD camera imaging at 80 fps which cues mirrors to direct light into a narrow-field telescopic system. Details of the hardware, calibration and reduction examples can be found in \citet{Weryk-CAMO-2013, Vida2021}; further details are in Tables~\ref{tab:CamDetails} and \ref{SpecTable}.

\begin{figure}
    \centering
    \includegraphics[width=0.75\textwidth]{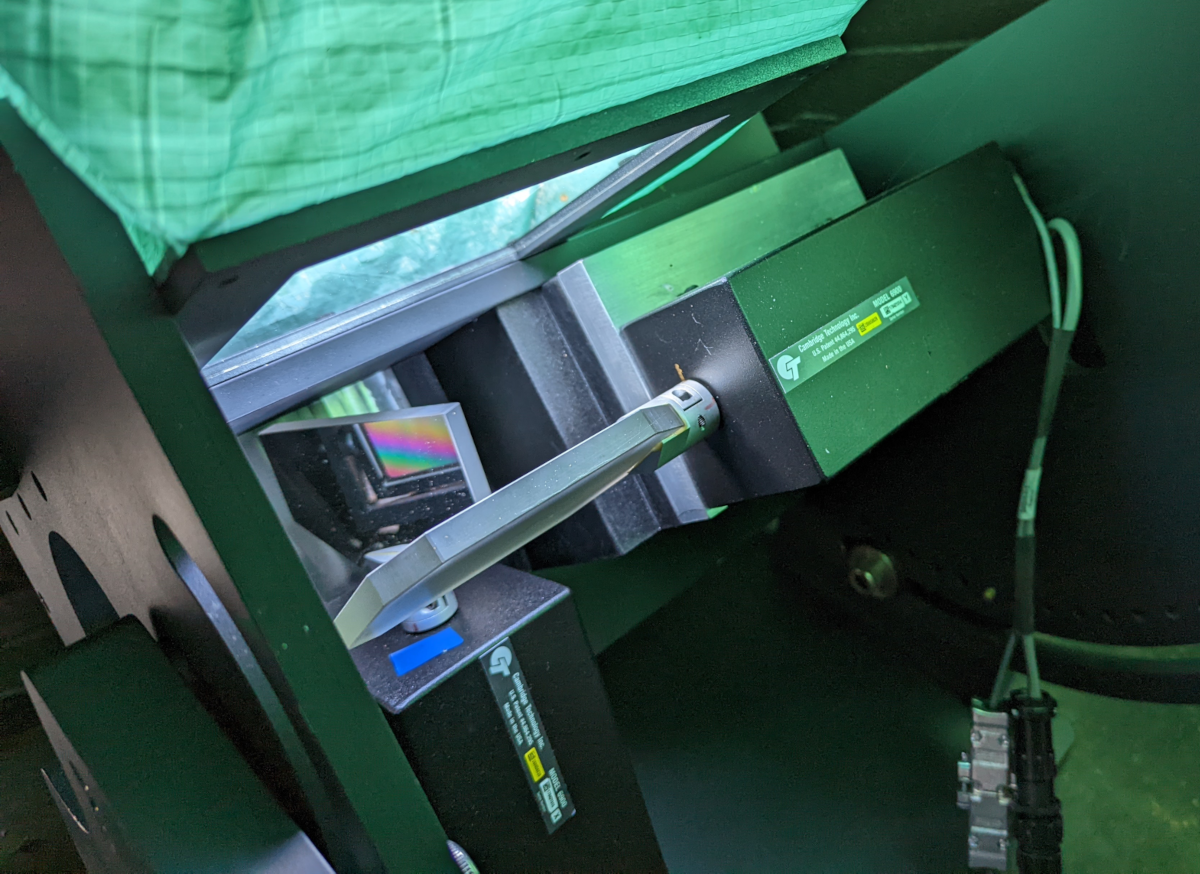}
    \caption{The two galvanometers for CAMO-S (center), housed in a climate-controlled enclosure, view the sky through a BK7 window (top). The light from the meteor reflects off the mirrors and to the left in this image where a grating (not shown) is placed together with the spectral camera.}
    \label{fig:CAMOinside}
\end{figure}

During clear, dark periods, meteors brighter than +3M are detected in real-time by the wide-field fixed camera and the meteor positions are transformed in real-time (latency less than 20 ms) to alt-azimuth coordinates. These are then converted to encoder positions to slew two orthogonal Cambridge Technologies model 6900 galvanometer mirrors with 50mm aperture that direct the light from the meteor to a telescope with an image intensified narrow-field camera at its focus. This is shown as setup in the field in Figure \ref{fig:CAMOinside}

The use of galvanometers in the tracking system not only allows for very rapid response times but also provides potential tracking speeds much greater than the apparent speed of even the fastest meteors. For a typical event, the system is able to acquire a meteor and begin tracking within a few tens of milliseconds after a track is established (which typically takes several initial frames of detection in the wide-field camera). The meteor typically drifts cross-track by less than a pixel between frames and in many cases less than this amount \citep{Stokan2013}.

The narrow-field tracking camera is capable of imaging meteors with peak magnitudes of +7 at frame rates of 110 fps and, due to the long focal length of the narrow-field telescope, resolutions of 70 cm/pixel at 100 km range. 

In 2020, the existing CAMO tracking system (hereafter CAMO-Optical or CAMO-O) at Elginfield was augmented with a coupled spectral system, CAMO-S (-S for spectral). The results presented here were collected after the 2023 upgrade of CAMO-S to use a more sensitive camera and less dispersive diffraction grating following initial engineering tests from 2020-2022. 

In the current mode of operation, meteors are detected by the wide-field camera of CAMO-O and the detection is used to steer mirrors for both CAMO-O and CAMO-S in parallel. Hence, most events are simultaneously captured in the CAMO-O mirror tracking system (provided the tracking quality is good) at both the Elginfield and Tavistock field sites. As a result, the complete trajectory and orbit are known for many CAMO-S meteors from the data gathered by CAMO-O. 

As a result, absolute elemental abundances can, in principle, also be computed. With the high precision of CAMO-O astrometry. In this configuration, CAMO-S facilitates simultaneous estimation of meteoroid composition to complement bulk density, mass, orbit, and fragmentation behaviour, including, for some cases, compositional variations among fragments as imaged with CAMO-O.

CAMO-S makes use of the same basic architecture as CAMO-O. Specifically, for CAMO-S, a second set of identical galvanometers and mirrors are positioned next to the existing narrow field system. 

In the CAMO-S beamline, before entering a long-focal-length lens for collimation, light is directed through a blazed transmission grating. After experimentation, a 150 lines per mm grating  was selected for permanent operation together with a low-noise Nuvu EMCCD camera \footnote{https://www.nuvucameras.com/products/hnu-1024/}. In early engineering tests, a Qimaging Extended Blue intensified camera \citep{Kikwaya2011} was chosen for its sensitivity in the blue to help identify lines in the 300-400 nm range. However, real-world testing demonstrated that the EMCCD has nearly the same spectral sensitivity but much better noise characteristics and it was adopted as the final CAMO-S spectral camera following commissioning for operations in 2023.  

\begin{figure}[h]
    \centering
    \includegraphics[width=\linewidth]{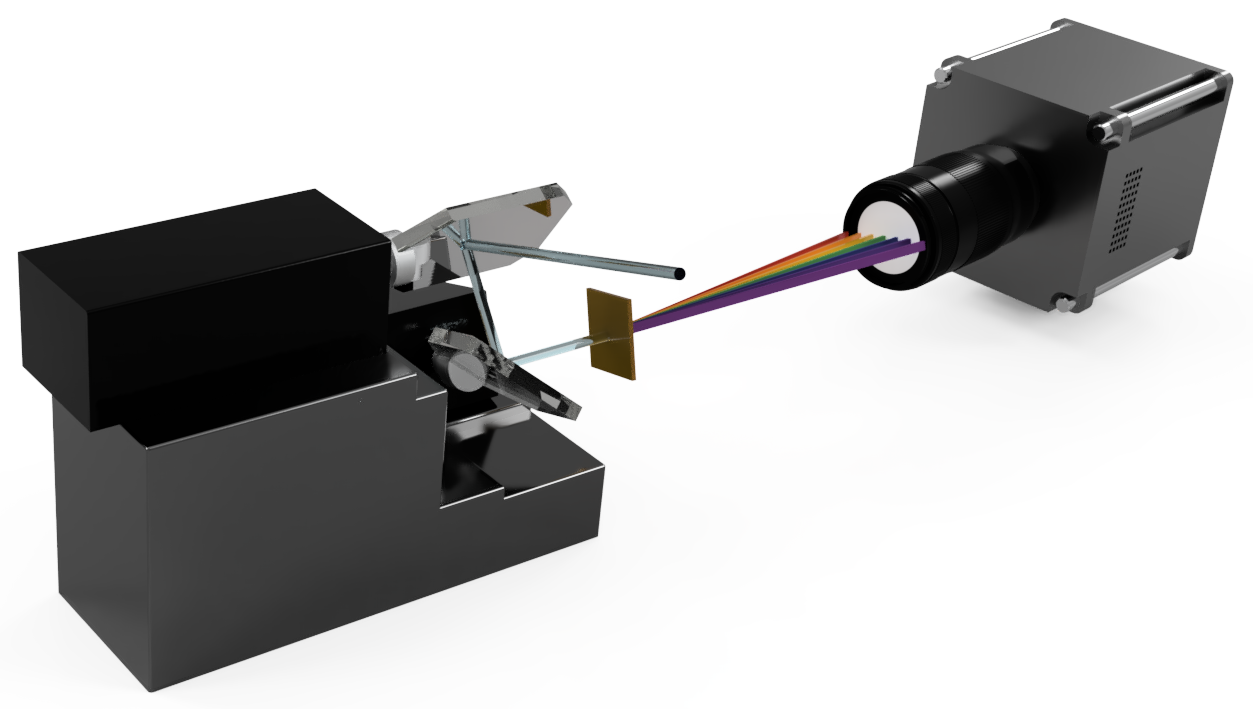}
    \caption{A rendering of the CAMO-S system which consists of two moving-coil galvanometers (black boxes on left at right angles to each other) with 50mm mirrors that direct the light from a tracked meteor through a transmission grating before being imaged by an EMCCD camera attached to a 180 mm photographic lens.}
    \label{fig:CamoSSchematic}
\end{figure}

The 1\textsuperscript{st}-order spectrum from the grating is imaged by the spectral camera at a frame rate of 32 fps. To achieve 32 fps, however, the camera needs to be run in 2x2 binned mode. This results in a 2x increase in frame rate, at the expense of spectral resolution, allowing us to better capture temporal changes in the spectra of short-lived meteors. The measured spectral scale of this system is 0.985 nm/pixel (binned 2x2). A second camera images the 0\textsuperscript{th}-order direct meteor view at 20-40 fps with a scale of 7.9"/pixel (Fig.~\ref{fig:CamoSSchematic}). This direct camera will eventually be used to automatically find the position of spectral lines in the spectral camera, but for this study, the lines were manually identified.

\begin{figure}[h]
    \centering
    \includegraphics[width=0.5\textwidth]{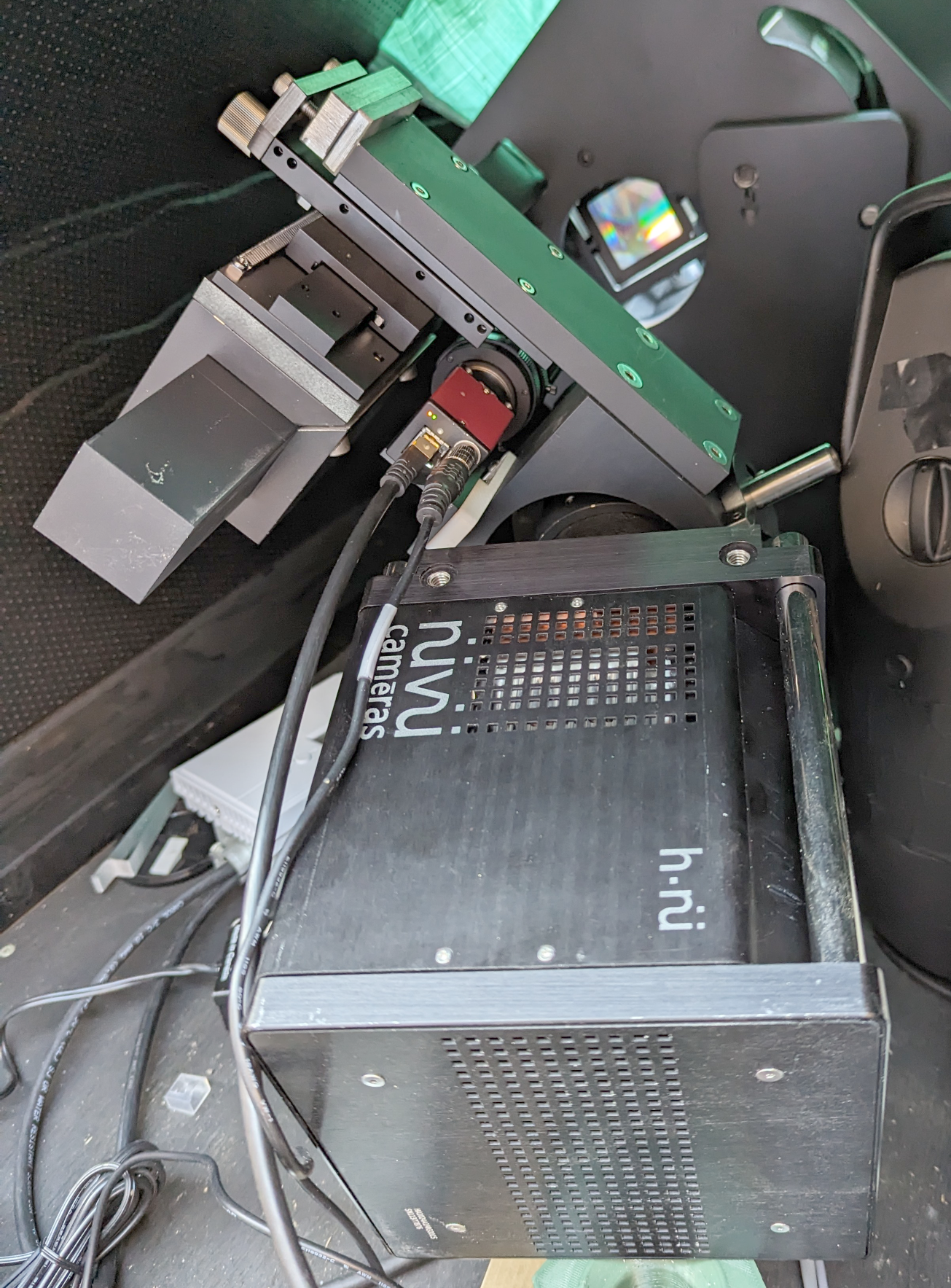}
    \caption{The CAMO-S system showing the EMCCD spectral camera (bottom) and direct camera (red camera at top). The grating can be seen at the top center through the opening in the galvanometer mount. }
    \label{fig:CAMO-S_EMCCD}
\end{figure}

\begin{table}
 \begin{tabular}{lcccc}
    \hline Camera & Resolution & Spectral Range (nm) & Frame Rate (fps) & Scale ($''$/pixel)  \\
    \hline
    Wide Field   & 900$\times$900 & 350-900  & 80 & 61 \\
    Narrow Field & 900$\times$900      & 350-900 & 100 & 1.3 \\
    Spectral Tracking & 1024$\times$1024 & 350-1100 & 32 (2$\times$2 binning) & 8.7 \\
    Spectral Direct & 1456$\times$1088  & 350-1100  & 75 & 7.9 \\
 \end{tabular}
    \caption{Camera details for CAMO-Optical and CAMO-S.}
    \label{tab:CamDetails}
    
\end{table}

\begin{longtable}{p{.40\linewidth} p{.60\linewidth}}
\caption{CAMO mirror tracking and spectral system hardware specifications.\label{SpecTable}}\\
\hline
 & Specification \\
\hline

\textbf{Wide-field Fixed Camera} \\
Camera & AVT GX1050 KAI-01050 (14-bit) (CCD)\\
Intensifier & Gen3 ITT 18-mm FS9910 \\
Lens & 25 mm f/0.85 \\

\textbf{Narrow-field Tracking Camera} \\
Camera & AVT GX1050 KAI-01050 (14-bit) (CCD)\\
Intensifier & Gen3 ITT 18-mm FS9910  \\
Telescope & Williams Optics Zenithstar 80-mm f/6.8 \\

\textbf{Spectral Tracking Camera (2023-onwards)} \\
Camera & Nuvu HN\"{u}1024 EMCCD (15-bit)\\
Lens & Nikon 180 mm f/2.8 \\
Grating & 44-mm transmission grating, 150 lpmm, 8.6\textdegree\  blaze angle (2023 onwards)
\\
\textbf{Spectral Direct Camera} \\
Camera & AVT Mako G-158 IMX273 (CMOS)\\
Lens & Kowa 9-90 mm f/1.8 \\
\\
\end{longtable}

\section{Data Reduction}

Within the CAMO-S system, meteor detection and data collection are automated procedures. During a night of observations, the CAMO-S EMCCD spectral camera runs in free run mode filling a video buffer during times when the CAMO-O system is also operating. Video data are subsequently cut out of the buffer based on the timing of event triggers from the wide-field camera. CAMO-S is thus using the same detection algorithm/pipeline as the original CAMO-O tracking system \citep{Weryk-CAMO-2013, Vida2021}. 

Basic data reduction involves loading the spectral video sequences and calibrating the pixel to wavelength scale on visible lines. The spectral responsivity is calibrated using a star with known spectrum; this can be a function of the location in the field of view. These steps produce a corrected spectrum in which elements may be identified. Details of the corrections applied to arrive at spatial and intensity calibrated spectra are given in Appendix~\ref{app:SpectralSensitivity}.

Individual meteor spectra are manually analysed and processed with a Python-based software package that we call Meteor Elemental Spectra Software (MESS). MESS is intended to be a complete visualization, calibration and modeling package for both tracked and untracked meteor spectra observations. It has been built around functions from the CAMS Spectral \citep{jenniskens2013CAMSS} C-code spectral library (written by P. Gural) and is currently being tested with data from CAMO-S. While MESS is designed to model the abundance of each element, for the current study we focus only on its visualization and calibration capabilities. A future paper will describe the modelling approach and results in more detail. 

\section{Observations and Initial Results}

Between August 2023 and May 2024, CAMO-S has had 3834 detections, of which 1364 were well-tracked and bright enough to produce useful spectra. Of these useful spectra, 600 were tracked from both stations and had trajectories and orbits. From this dataset we estimate that CAMO-S can produce useable spectra for meteors brighter than magnitude +2, irrespective of angular velocity (as CAMO-S tracks each event). As an example of the final data product, we have selected some of the more interesting events captured during initial instrument commissioning which also address our initial design goals. 

The spectral system ran in its final configuration, with the EMCCD camera and 150 lpmm diffraction grating and alignment corrected beginning in the summer of 2023. 

\begin{figure}
    \centering
    \includegraphics[width=0.9\linewidth]{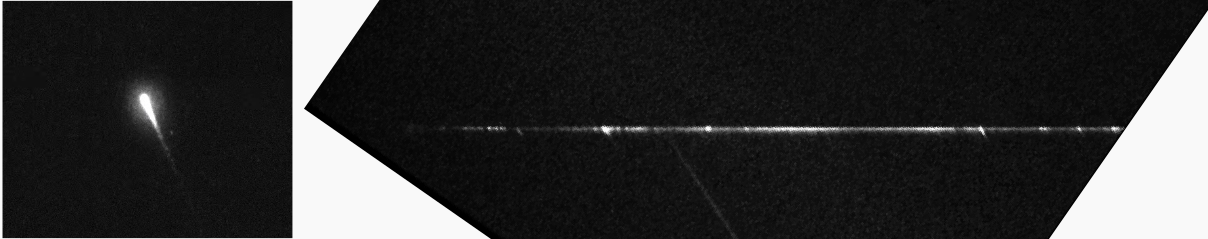}
    \caption{Single-frame image cutouts of a meteor imaged by both the Elginfield CAMO optical (left) and spectral (right) systems on August 5, 2023. A 55.2\textdegree{} rotation has been applied to the spectral image.}
    \label{fig:20230805_025924-Comparison}
\end{figure}

\begin{figure}
    \centering
    \includegraphics[width=0.75\textwidth]{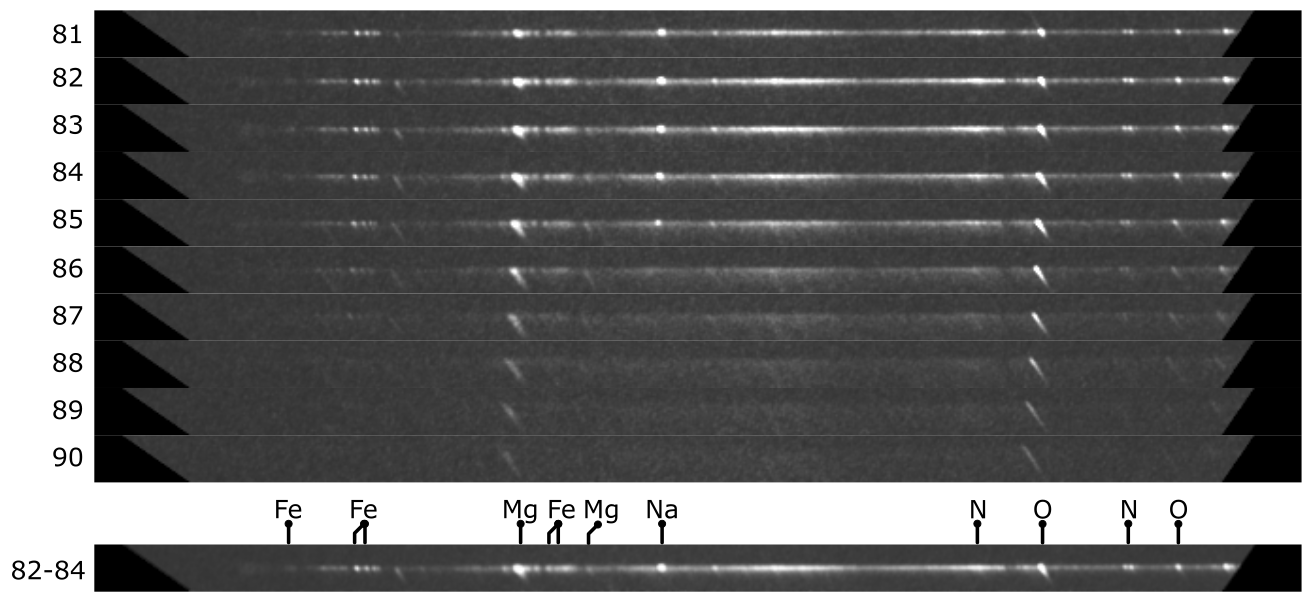}
    \caption{A 32.5 km/s meteoroid observed on August 5, 2023 which shows numerous meteoric metal lines. The 150 lpmm grating used here gives spectral scale of 0.935 nm/pixel. The detectable wavelength range is from about 380 nm to more than 850 nm. The y-axis numbering indicates frame numbers and the final row is a combined stack of the 3 brightest frames. Note that the camera is running at 32 frames per second. The spectrum is first detected at a height of 98 km and the last frame with detectable spectra is at a height of 90 km. The meteor reached a peak apparent magnitude of +0.6. The meteoroid had a pre-atmospheric orbit of a=3.2 AU, e=0.7, i=51.2$^\circ$, Q=5.4 AU and a geocentric radiant of $\alpha_g$=302.1$^\circ$, $\delta_g$=63.3$^\circ$. }
    \label{fig:EMCCDtracked_20230805-025924}
\end{figure}

\begin{figure}
    \centering
    \includegraphics[width=0.75\textwidth]{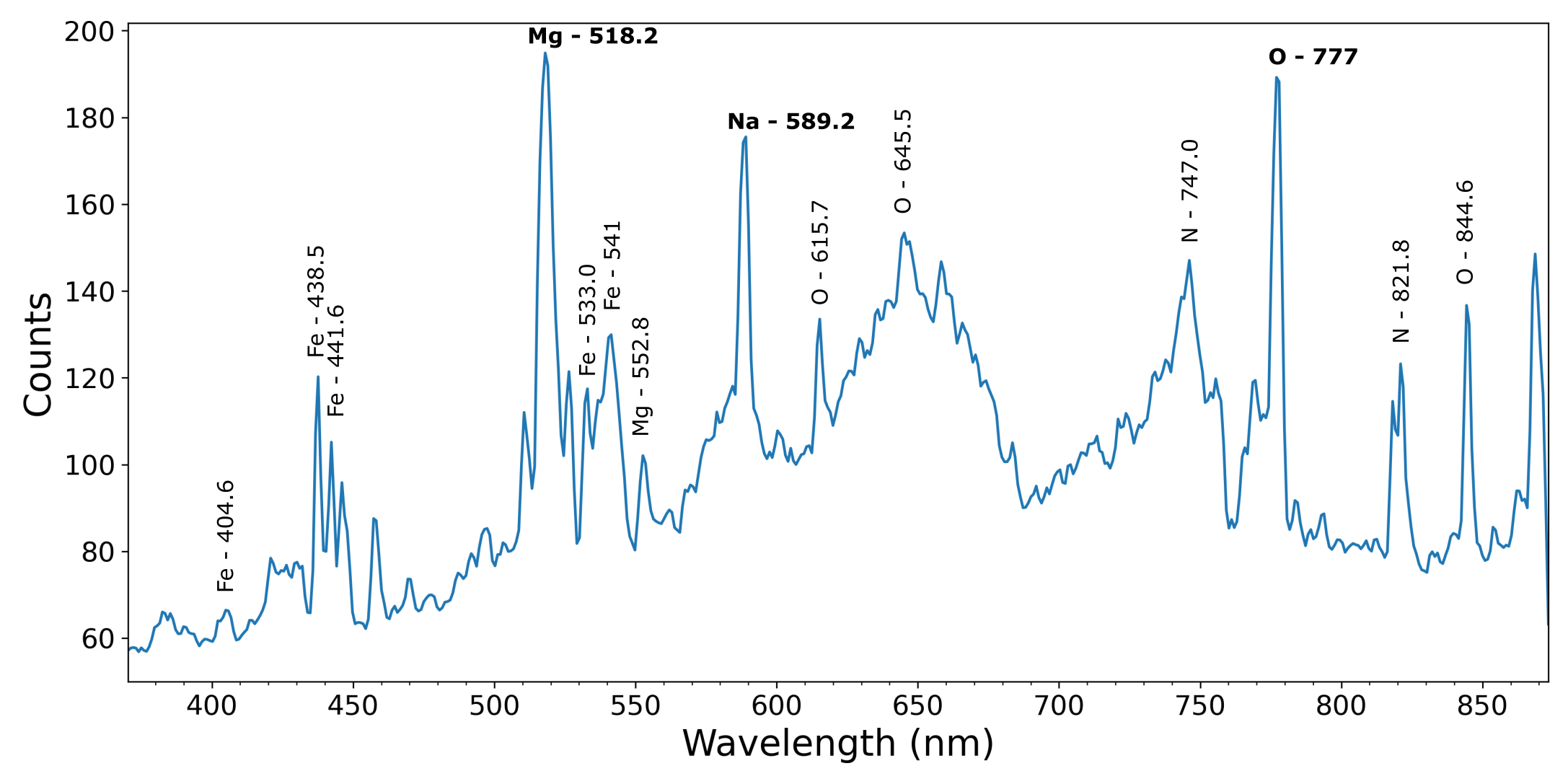}
    \caption{Line plot of the  meteor shown in Figure 3. Here individual lines are identified based on their expected strength (from MESS) and position.}
    \label{fig:specplot_20230805-025924}
\end{figure}

 An example of a typical early result is shown in Figure \ref{fig:20230805_025924-Comparison}. These show single sub-frame cutouts from CAMO-O and CAMO-S data captured at the same time for a common meteor while Figures \ref{fig:EMCCDtracked_20230805-025924} and \ref{fig:specplot_20230805-025924} provide the details of the time evolution of the spectra and detailed line identification, respectively. Together with hundreds of other tracked events, these initial observations indicate that meteors were generally well-tracked and centred in the camera field of view, a central requirement for the system. A linear dispersion of 0.935 nm per pixel was derived, giving a possible spectrum range of 585 nm when the camera is rotated at an angle of -55.2 degrees (relative to the sensor's horizontal axis). This configuration allows the spectrum to fall diagonally on the sensor, maximizing the length of the spectrum when the meteor is well-tracked. 

\subsection{The August 2023 Perseid meteor shower}

\begin{figure}
    \centering
    \includegraphics[width=0.9\textwidth]{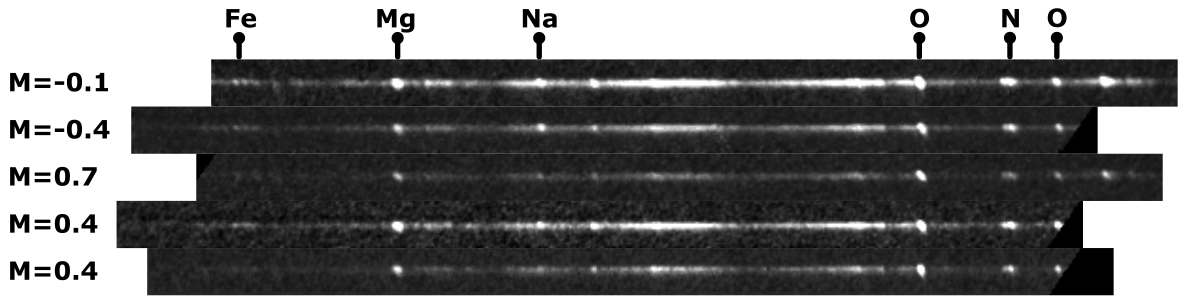}
    \caption{A collection of 5 Perseid spectra observed with CAMO-S on August 11, 2023. Times of the peak brightness in UTC are given. The major elements identified are indicated.}
    \label{fig:Perseids_20230811}
\end{figure}

Shortly after initial equipment commissioning with the system in its final configuration with the EMCCD camera and the 150 lpmm grating we had an opportunity to further test the system during the 2023 Perseids. The night of August 11, 2023 was mostly cloudy; however the system ran for about 5 hours and collected 28 spectra, 5 of which were Perseids with tracking quality sufficient for interpretation (Figure \ref{fig:Perseids_20230811}).

Using the Perseid spectra the spectral scale for the EMCCD with the 150 lpmm grating was confirmed. Each spectrum was scaled and rotated after which the pixel positions of the Mg (518.2 nm) and O (777.1 nm) lines were measured. The average of all measurements indicated a spectral scale of 0.934 \textpm 0.002 nm per pixel. Although we assume, for the sake of simplicity, that the spectral scale is constant across the sensor and throughout the spectral range of the system, this is likely not strictly the case operationally. Future work will examine and quantify this effect, but, for the purposes of this initial study, this approximation is sufficient.

\begin{figure}
    \centering
    \includegraphics[width=0.9\textwidth]{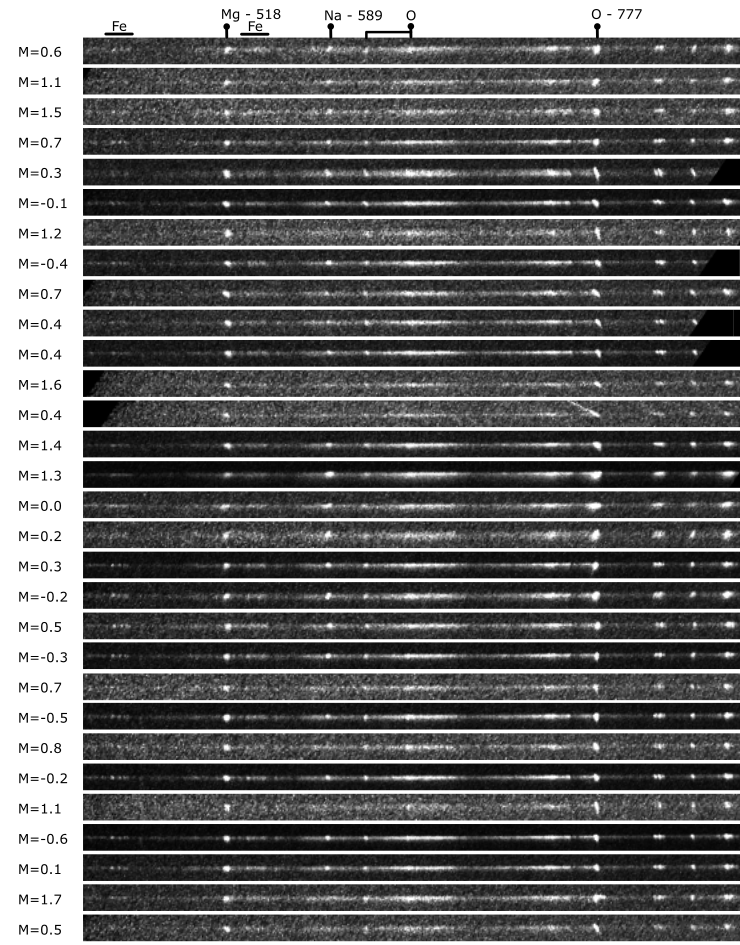}
    \caption{A compendium of 30 2-station Perseids imaged by Camo-S (and CAMO-O) during the 2023 Perseid meteor shower. The events range in peak magnitude from 1.7 to -0.6. All spectra are uncorrected for flat field but have been auto-scaled.}
    \label{fig:Perseids_2023-1}
\end{figure}

\begin{figure}
    \centering
    \includegraphics[width=0.9\textwidth]{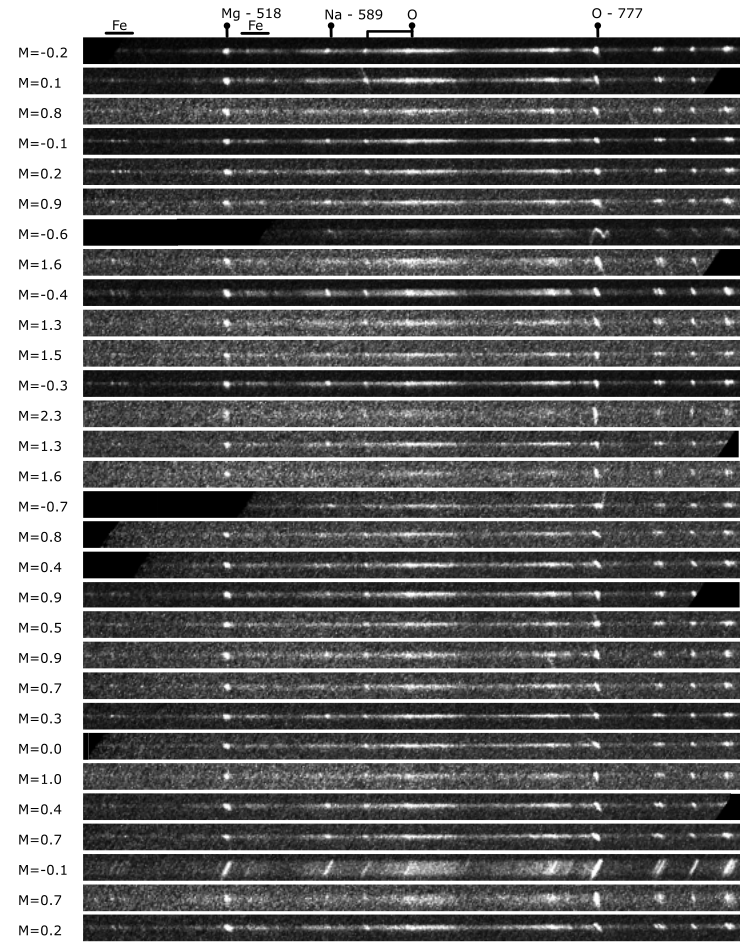}
    \caption{A second compendium of an additional 30 2-station Perseids imaged by Camo-S during the 2023 Perseid meteor shower. These events range in peak magnitude from 2.3 to -0.7. All spectra are uncorrected for flat field but have been auto-scaled.}
    \label{fig:Perseids_2023-2}
\end{figure}

In total, 60 Perseid spectra ranging in magnitude from -0.7 to 2.3 (Figures \ref{fig:Perseids_2023-1} and \ref{fig:Perseids_2023-2}) were collected during the entire 2023 Perseid meteor shower. As expected, the spectra are similar to one another with the main difference being in the relative strength of sodium \citep{Matlovic2019}. 

The reliable identification of lines and the variations in their relative strengths demonstrates that one of the goals of CAMO-S, namely to provide relative strengths of the major elemental lines, particularly Fe, Na, and Mg has been achieved.

\subsection{The visibility of the Ca(II) H and K lines}

As one of the primary goals of this project is to identify the presence of refractory inclusions, CAMO-S should be able to resolve the singly-ionized Ca(II) H and K lines at 393.4 nm and 396.8 nm. Although our modeled responsivity of the system suggested a spectral range from about 375 nm to roughly 900 nm, we have found that, in practice, the response below 400 nm is lower than initially thought. As a result, few meteors showing obvious Ca(II) H and K emission lines have been observed to date. Given that the H and K lines have high excitation temperatures - on the order of 10,000 K - this result is maybe not too surprising as the hottest/densest plasmas would be expected for fast meteors and fireballs, not with the fainter, more moderate speed meteoroids that we typically observe with CAMO-S. Even so, several meteors showing bright Ca(II) H and K lines have been recorded.

\begin{figure}
    \centering
    \includegraphics[width=0.9\textwidth]{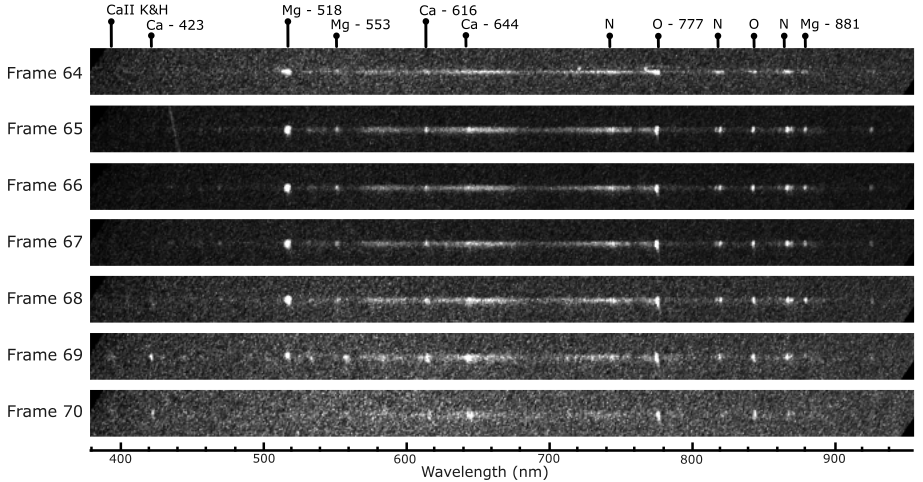}
    \caption{A meteor captured on June 1, 2024 at 4:10:37 UT shows neutral Ca lines as well as the Ca(II) H and K lines. This 2-station meteor had a measured speed of 46 km/s and a peak brightness of M=0.5.}
    \label{fig:20240601_041037}
\end{figure}

The first example (Figure \ref{fig:20240601_041037} we present was observed June 1, 2024 at 4:10:37 UT with CAMO-S and both CAMO optical tracking cameras. The automated two-station solution suggests that this was a 100 mg meteoroid travelling at 46.0 km/s with a peak brightness of M=0.5. The orbit was fairly eccentric (e=0.6) with an inclination of 83\textdegree{}  and a q\textsubscript{per} of 0.96 AU. The begin and end heights are 92.7 km and 85.5 km, respectively. With its moderate speed and relatively short duration (0.25 s), this meteor was observed in just seven spectral image frames. Initially, the spectrum shows a strong Mg line as well as atmospheric O and N lines. Both Fe and Na appear to be absent, but there are indications that Ca and K may be present in small amounts. By the third frame, the SNR of the spectrum has improved and the Mg, Ca and atmospheric lines are all distinct. The fifth and sixth frames shows diminished Mg line strength and increased Ca. Ca(II) H and K lines first become identifiable in the sixth frame. In the final frame, the Mg line has disappeared while faint Ca(II) H and K lines along with Ca lines persist. 

\begin{figure}
    \centering
    \includegraphics[width=0.9\textwidth]{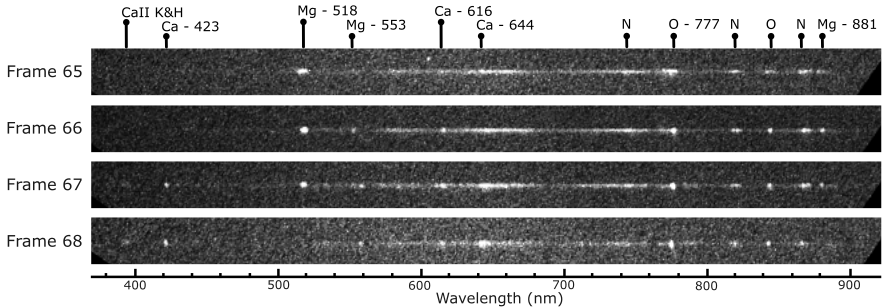}
    \caption{A second two-station meteor showing neutral Ca and Ca(II) H and K lines was observed June 1, 2024 at 7:57:22 UT. This meteor had a speed of 52.7 km/s and a peak magnitude of M=1.1.}
    \label{fig:20240601_075722}
\end{figure}

The second example (Figure \ref{fig:20240601_075722}) that we present was captured on June 1, 2024 at 7:57:22 UT. Similar to the previous example, the two-station solution suggests a roughly 100 mg mass meteoroid with a speed of 52.7 km/s and a peak brightness of M=1.1. The orbit of this meteoroid was highly eccentric (e=0.94) with a high inclination of 92.6\textdegree{} and a q\textsubscript{per} of 0.75. Begin and end heights are 94.5 km and 86.4 km, respectively. Due to the short duration (0.2 s), this solution has large uncertainties and should be interpreted accordingly. The first two frames of data show a spectrum containing Mg and atmospheric N and O lines, with no indications of Na or Fe. Neutral Ca and Ca(II) lines are faint or absent for these first frames. The last two frames of this sequence, however, show the appearance of the Ca(II) H and K lines along with Ca lines at 423 nm, 616 nm, and 644 nm. Here, Ca and Ca(II) dominate with the only other distinct lines coming from atmospheric N and O.

While identification of Ca lines is much less common with CAMO-S than originally hoped, the system does clearly probe Ca composition, a key design goal. Improved Ca sensitivity is part of planned future improvement of the system. 

\subsection{The spectra and fragmentation behaviour of iron meteoroids}

\begin{figure}
    \centering
    \includegraphics[width=0.9\textwidth]{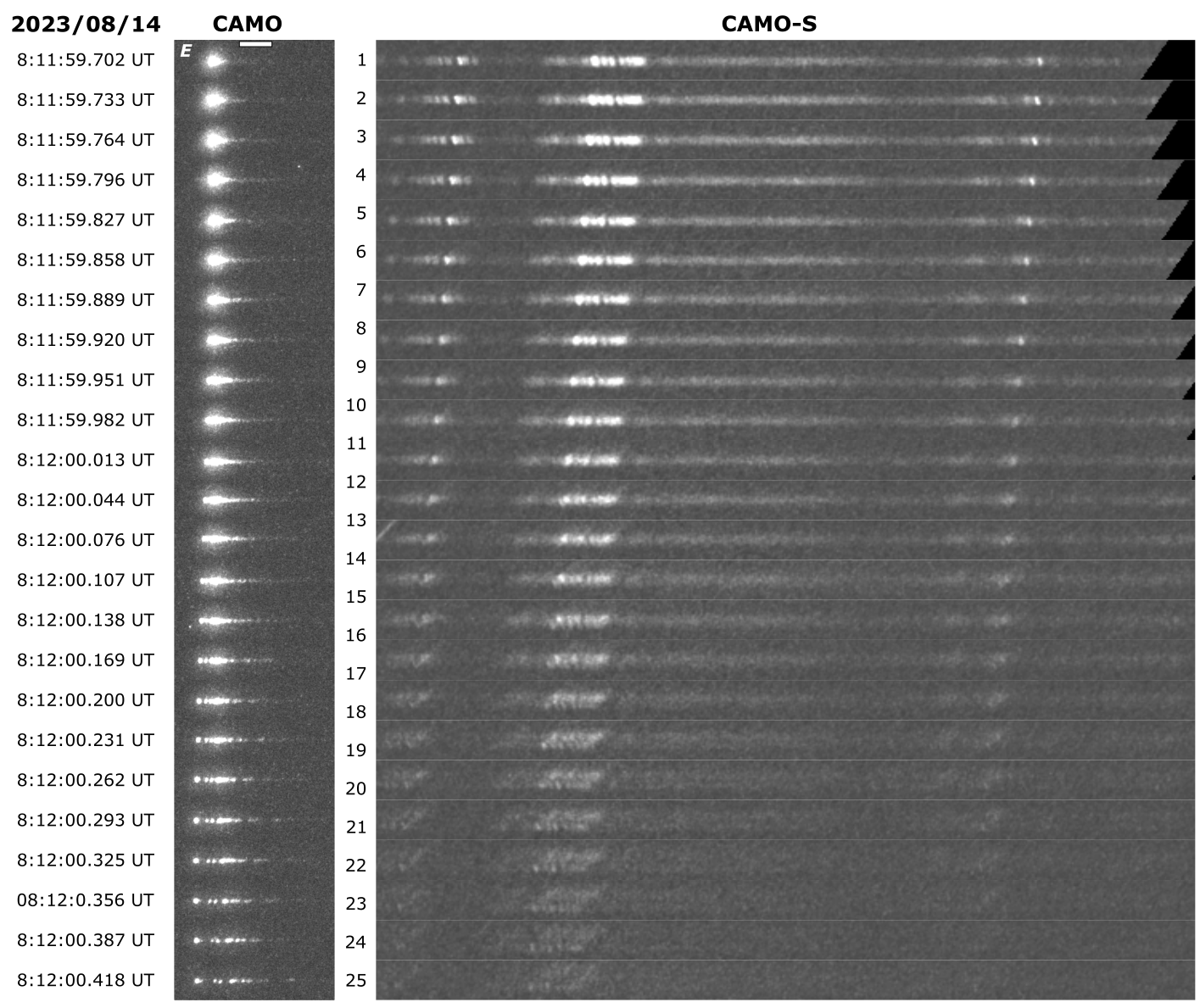}
    \caption{A single station fragmenting iron meteoroid observed on August 14, 2023 at 8:12:00 UT by both CAMO optical and spectral tracking systems. Approximately a dozen fragments are visible in the optical imagery while 2-3 separated spectra of fragments are observed in the spectral imagery in the final few frames. Note the scale bar at the top of the CAMO-O image stack represents 40 m and this imagery is from Elginfield (E).}
    \label{fig:IronMeteoroid_CamosComparison}
\end{figure}

\begin{figure}
    \centering
    \includegraphics[width=0.9\textwidth]{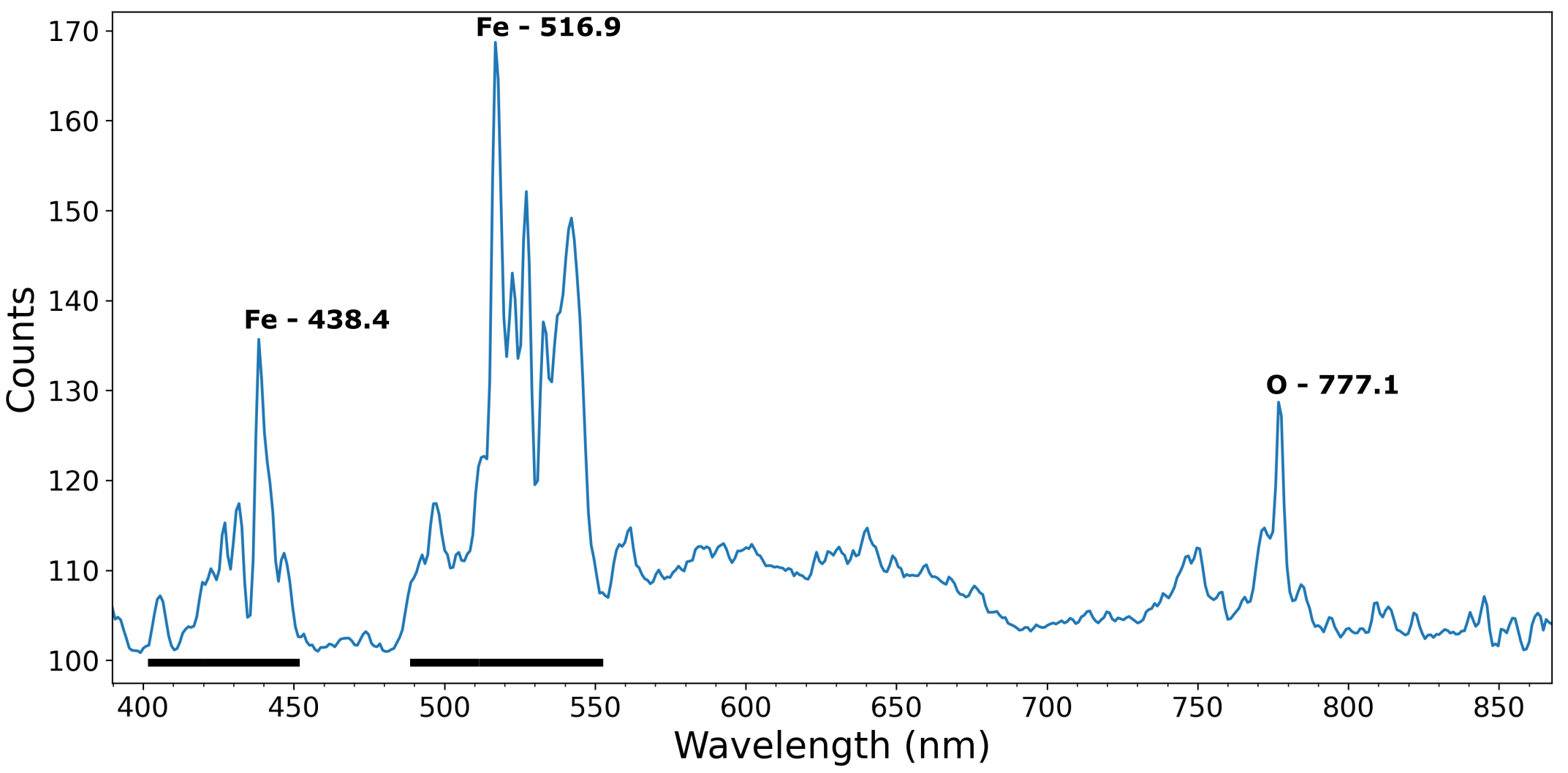}
    \caption{A line profile of the spectrum of the iron meteoroid observed by CAMO-S on August 14, 2023 at 8:12:00 UT as shown in Figure \ref{fig:IronMeteoroid_CamosComparison}. The abundant iron lines (marked by the black bands) are the dominant feature.}
    \label{fig:specplot_20230814-08120}
\end{figure}

On August 14, 2023 at 08:12:00 UT a +2 magnitude meteor was tracked and imaged by both the CAMO optical and spectral systems at Elginfield. The optical tracker showed a bright meteoroid disintegrating into at least a dozen fragments during its passage through the atmosphere (Figure \ref{fig:IronMeteoroid_CamosComparison}). While this is not uncommon, this event was the first fragmenting meteor that we investigated in detail with the spectral system.  The observed spectrum suggests that the object was predominantly composed of iron (Figure \ref{fig:specplot_20230814-08120}) while weak atmospheric lines suggest a moderate speed of entry.

From the optical imagery, we see that the meteor exhibits signs of fragmentation throughout its observed trajectory. Beginning at frame 7, the meteoroid begins to disintegrate into large fragments which, on their own, appear to be non-fragmenting as they continue to further ablate. In all, about a dozen individual fragments can be observed and tracked. We suspect that these are not solid fragments but more likely iron molten droplets as previously suggested by \citet{Capek2019}.

In terms of the speed of this particular meteoroid, we only have single-station observations so can not determine the trajectory and orbit. However, the presence and moderate strength of the atmospheric oxygen line at 777.1 nm would suggest a low-to-moderate speed - likely in the range of 20-30 km/s. 

As stated earlier, one of the goals of this project was to be able to image the spectra of individual fragments from a disintegrating meteoroid. A close examination of Figure \ref{fig:IronMeteoroid_CamosComparison}  reveals that, at frame 15 in the spectral data, a splitting of the spectrum begins to appear -- note that the spatial resolution of the spectral system is lower, so the fragments are closer together than on CAMO-O images. As the distance between the largest, most luminous fragments increases, so too does the separation of the fragment spectra. And, by the end of the meteoroid's flight, there appear to be three distinct, albeit faint, spectra. This implies gross fragmentation as opposed to the quasi-continuous fragmentation commonly believed to occur with iron meteoroids (\cite{Ceplecha1966, Ceplecha1967, Bronshten1981}). In this case, the spectra of all three fragments appear to be the same and all are dominated by iron.

With the successful optical and spectral imaging of the August 14, 2023 event, six months of data collected between August 2023 and February 2024 was manually searched for additional iron meteoroids. In addition to strong iron lines in the spectra, we also noted a number of spectra that showed a medium-strength sodium line in addition to strong iron lines. Compared to the work of \citet{Borovicka2005}, what we refer to as iron-rich spectra possibly correspond to their `enhanced-Na' group. Of the more than 850 spectra collected during this period, 44 were identified as coming from iron meteoroids while a further 17 were considered to be iron-rich (summarized in Table \ref{table:IronMeteoroids}). A more precise classification of these spectra using ternary plots will be done in future studies. 

A total of 63 primarily iron meteoroids are summarized in Table \ref{table:IronMeteoroids}; 22 of these have two-station data with computed trajectories and orbits. Those which are marked "Fe-rich" have other lines like Na in addition to strong iron lines. Six of the meteors have multiple fragments in which spectra from individual fragments can be distinguished; several of these are described below. 

\begin{figure}
    \centering
    \includegraphics[width=0.9\textwidth]{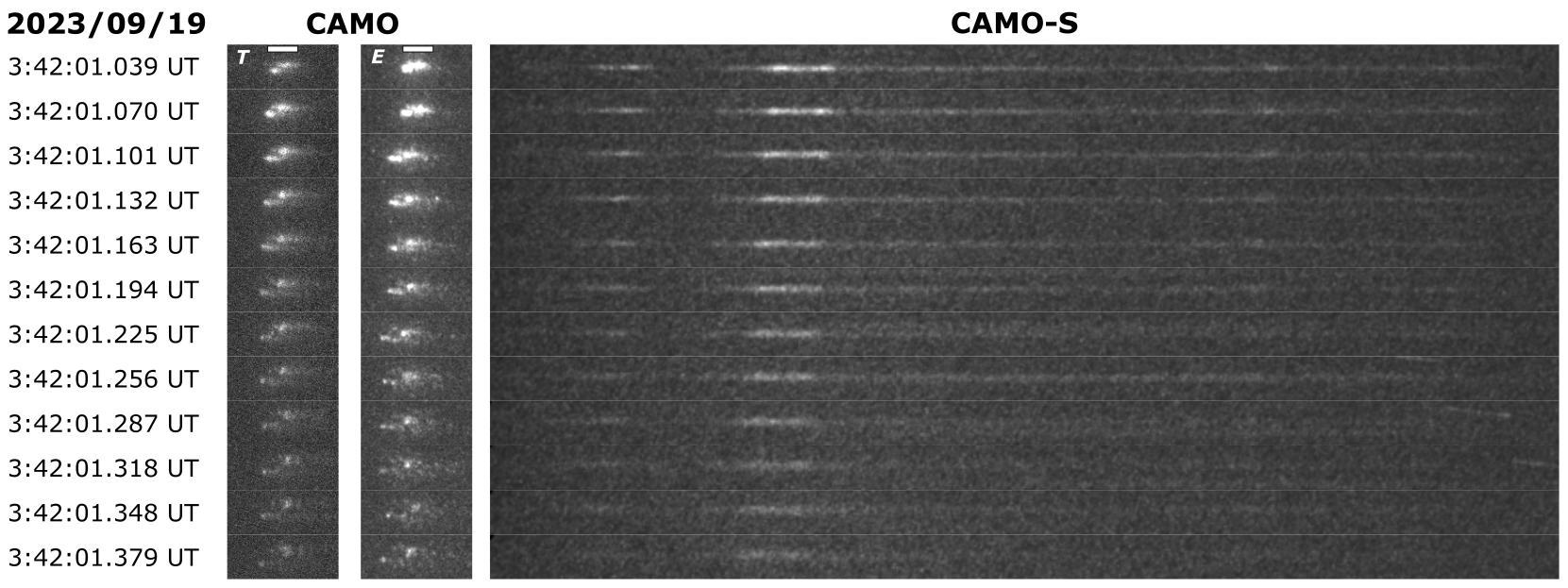}
    \caption{An iron meteoroid occuring on Sep 19, 2023 showing transverse fragmentation observed from both CAMO sites. A two-station solution derived from CAMO tracking cameras at Elginfield (E) and Tavistock (T) gives v$_g$=21.5 km/s and a peak magnitude of +2.1. The scale bars in both CAMO-O stacks represent 40 m at Elginfield and 40 m at Tavistock.}
    \label{fig:IronMeteoroid_20230919-034201}
\end{figure}

The second iron meteoroid examined in detail was observed on September 19, 2023 at 3:42:01 UT by both CAMO optical stations as well as CAMO-S (Figure \ref{fig:IronMeteoroid_20230919-034201}). The two-station solution gives a v$_g$ of 21.5 km/s, a peak magnitude of +2.1, and a derived mass of 6.8$\times$ $10^{-5}$ kg (Table \ref{table:IronMeteoroids}). From the optical cameras we see early transverse (perpendicular to direction of travel) fragmentation with several large fragments separated by 10-50 m. When aligned in time with the optical imagery, the spectral camera shows a spectrum comprised of mostly iron lines (multiplets 42, 41, and 15). In addition to iron, atmospheric oxygen and molecular nitrogen bands are also visible. In several of the spectral frames individual spectra for the two brightest fragments is observed.

\begin{figure}
    \centering
    \includegraphics[width=0.9\textwidth]{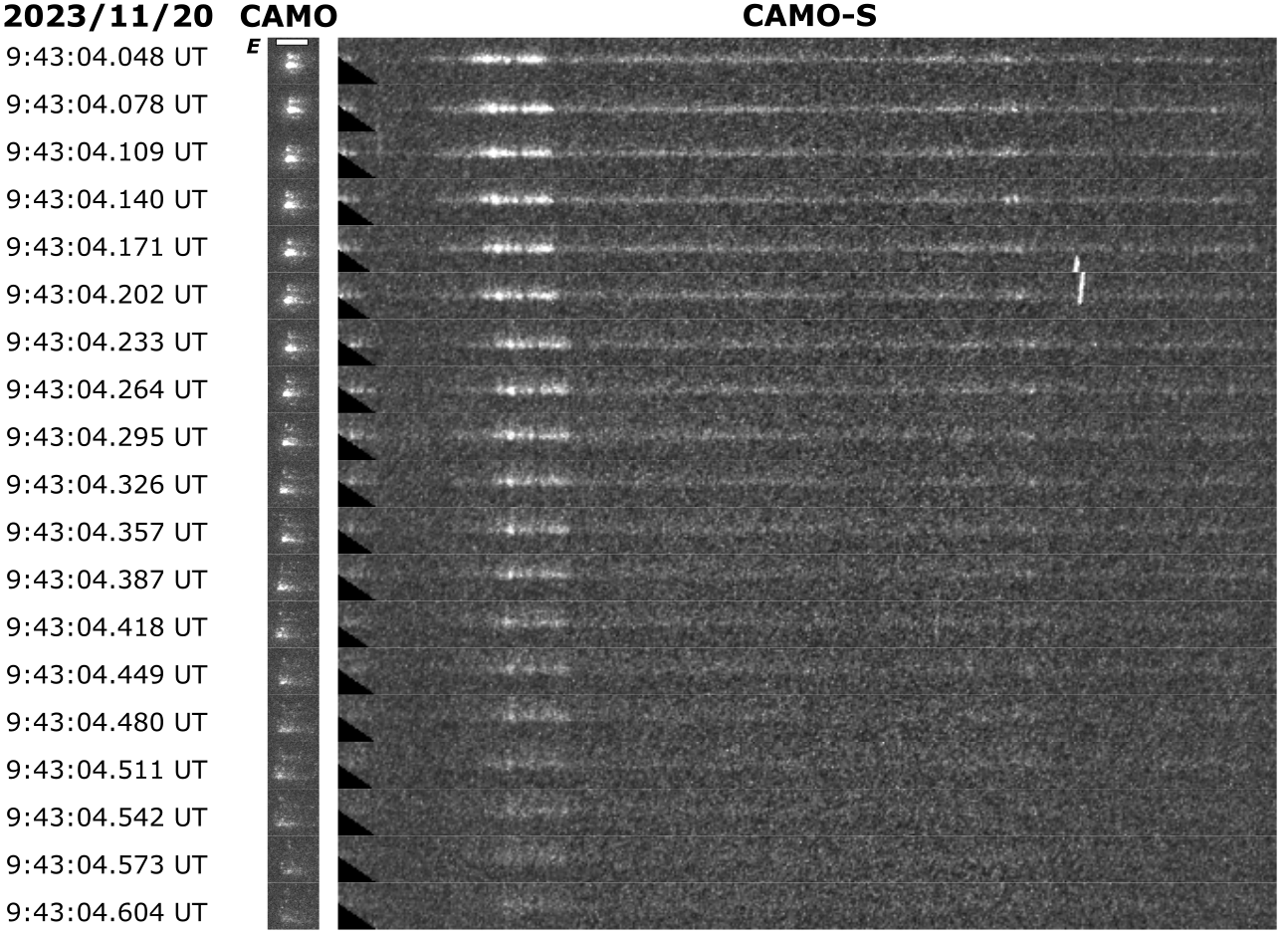}
    \caption{The fragmentation of a meteoroid tracked on Nov 20, 2023  shows a roughly 20 m transverse separation of the largest fragments. }
    \label{fig:IronMeteoroid_20231120-094304}
\end{figure}

Another transversely fragmenting meteoroid was observed on November 11, 2023 at 9:43:04 UT (Figure \ref{fig:IronMeteoroid_20231120-094304}). Since it was only observed from the Elginfield site, no orbit solution is available. Even without two-station data, however, we note the similarity between this event and the September 19, 2023 event. Iron lines between 517 nm and 545 nm are clearly visible as are atmospheric nitrogen and oxygen. Although we don't have a direct measure of the speed, we estimate it to be similar to the September 19, 2023 event as the observed atmospheric oxygen lines are of a similar strength. With a transverse separation of about 20 m and a favourable position angle relative to the direction of the grating ruling, the spectra from the two largest fragments show a separation sufficient for basic interpretation.

\begin{figure}
    \centering
    \includegraphics[width=0.9\textwidth]{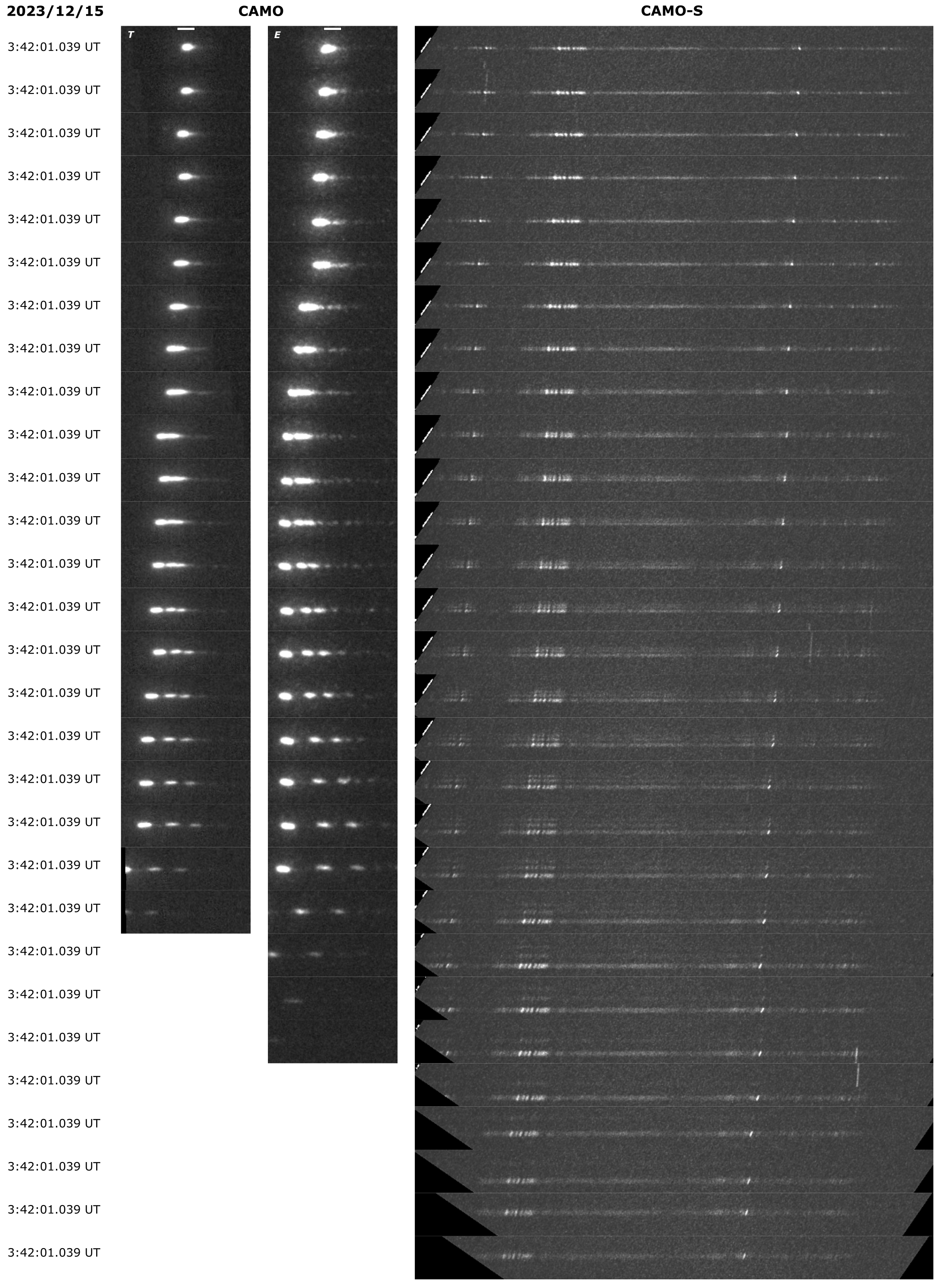}
    \caption{Inline fragmentation of an iron meteoroid which shows clear separation of the spectra from the three brightest fragments.The two-station solution gives a v$_g$=26.4 km/s and a peak magnitude of +2.1}
    \label{fig:IronMeteoroid_20231215-105317}
\end{figure}

On December 15, 2023 a bright (M$_{Peak}\approx +2$) was observed by all three optical and spectral tracking cameras (Figure \ref{fig:IronMeteoroid_20231215-105317}). As with our previous iron events, early frames from the optical cameras appear to show a quasi-continuous fragmentation. After the first few frames, however, gross fragmentation becomes apparent with large, bright fragments trailing one another at separations exceeding 100 m by the end of luminous flight. With these large separations, The spectral imagery shows distinct and interpretable spectra from the three brightest fragments. For each of these fragments, strong iron lines along with atmospheric nitrogen and oxygen are apparent. 

\begin{figure}
    \centering
    \includegraphics[width=0.9\textwidth]{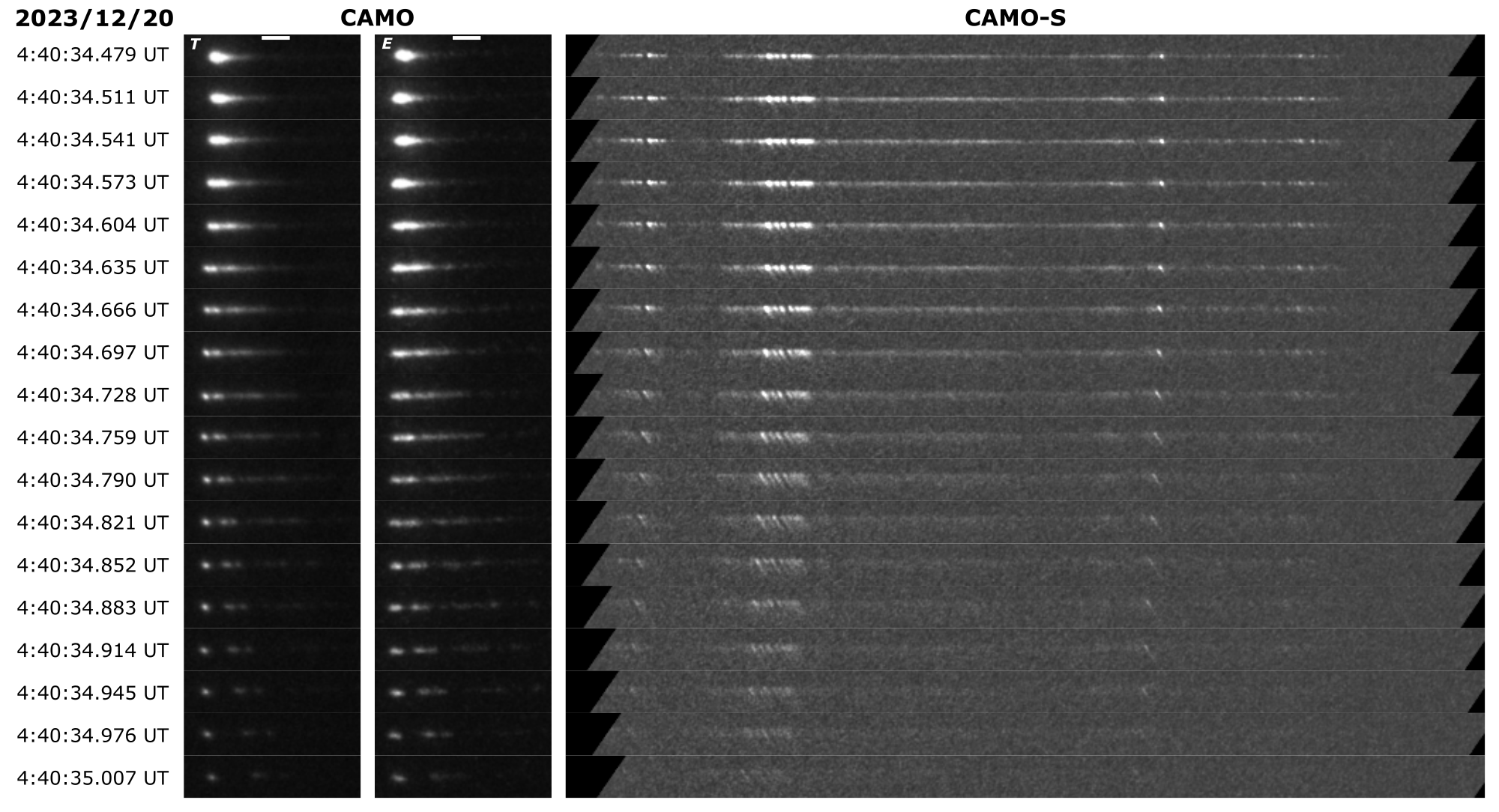}
    \caption{An iron meteoroid tracked on Dec 20, 2023 observed from both Tavistock and Elginfield. The trajectory solution gives a v$_g$=24.4 km/s and a peak magnitude of +1.5.}
    \label{fig:IronMeteoroid_20231220-044034}
\end{figure}

Finally, another bright meteor, with a peak magnitude of +1.5 and a speed of v$_g$=24.4 km/s, was observed on December 20, 2023 at 4:40:34 UT (Figure \ref{fig:IronMeteoroid_20231220-044034}). The optical cameras at the Elginfield and Tavistock stations both show gross fragmentation with fragment separations up to about 50 m, while the spectral camera shows strong iron lines as well as the atmospheric N$_2$ and oxygen lines expected for a meteor of this speed. Individual fragment spectra are visible, however, they are not as distinct as with the December 15, 2023 event.

In summary, the foregoing examples show that for meteors displaying gross-fragmentation and physical separations of a few tens of meters the current system is able to resolve spectra from discrete fragments. However, these spectra are often weak and the fraction of events meeting these criteria are small (of order a few percent of all CAMO-S events). Future modifications to the system will include a longer focal length imaging lenses to improve the detectability of individual fragment spectra.  

\begin{longrotatetable}
\begin{deluxetable}{c c c c c c c c c c c c c l}
\tabletypesize{\scriptsize}
\tablecaption{Likely iron and iron-rich meteoroids observed with CAMO-S.}
\label{table:IronMeteoroids}

\tablehead{
    \colhead{Date} & \colhead{Time} &
    \colhead{v$_{geo}$} & \colhead{M$_{peak}$} &
    \colhead{log(m)} & \colhead{h$_{beg}$} &
    \colhead{h$_{end}$} & \colhead{a} &
    \colhead{e} & \colhead{q} &
    \colhead{i} & \colhead{$\omega$} &
    \colhead{T$_J$} & \colhead{Notes} \\
    & \colhead{(UT)} & \colhead{(km/s)} &
    \colhead{(mag)} & \colhead{(kg)} & \colhead{(km)} &
    \colhead{(km)} & (au) & & (au) & \colhead{(\textdegree)} & \colhead{(\textdegree)} &
}
    
\startdata
        2023/08/06 & 03:33:59 & 21.2 & 0.5 & -3.32 & 100.4 & 88.9 & 2.51 & 0.71 & 0.73 & 2.4 & 251.0 & 3.05 & Fe-rich \\
        2023/08/09 & 02:44:22 & 12.8 & 1.4 &  & 111.8 & 87.8  & 1.22 & 0.25 & 0.92 & 1.3 & 238.7 & 5.19 & Fe-rich \\
        2023/08/11 & 04:59:54 & \nodata & 1.7 & \nodata & \nodata & \nodata & \nodata & \nodata & \nodata & \nodata & \nodata & \nodata & \\
        2023/08/11 & 06:17:44 & 22.0 & 1.6 & -3.80 & 98.3 & 89.6 & 2.67 & 0.73 & 0.72 & 6.9 & 252.5 & 2.9 & Fe-rich \\
        2023/08/13 & 04:19:08 & \nodata & 2.2 & \nodata & \nodata & \nodata & \nodata & \nodata & \nodata & \nodata & \nodata & \nodata & Fe-rich \\
        2023/08/13 & 05:59:11 &  \nodata & 2.9 & \nodata & \nodata & \nodata & \nodata & \nodata & \nodata & \nodata & \nodata & \nodata & Fe-rich \\
        \textbf{2023/08/14} & \textbf{08:11:59} & \nodata & 1.9 & \nodata & \nodata & \nodata & \nodata & \nodata & \nodata & \nodata & \nodata & \nodata & Distinct split spectra \\
        2023/08/16 & 09:06:38 & \nodata & 2.4 & \nodata & \nodata & \nodata & \nodata & \nodata & \nodata & \nodata & \nodata & \nodata & \\
        2023/08/17 & 01:48:40 & \nodata & 1.8 & \nodata & \nodata & \nodata & \nodata & \nodata & \nodata & \nodata & \nodata & \nodata & \\
        2023/08/19 & 04:23:11 & \nodata & 2.0 & \nodata & \nodata & \nodata & \nodata & \nodata & \nodata & \nodata & \nodata & \nodata & Fe-rich \\
        2023/08/19 & 04:25:17 & \nodata & 1.9 & \nodata & \nodata & \nodata & \nodata & \nodata & \nodata & \nodata & \nodata & \nodata & Fe-rich \\
        2023/08/20 & 05:10:17 & \nodata & 3.1 & \nodata & \nodata & \nodata & \nodata & \nodata & \nodata & \nodata & \nodata & \nodata & Possible NIA \\
        2023/08/20 & 07:27:56 & \nodata & 2.5 & \nodata & \nodata & \nodata & \nodata & \nodata & \nodata & \nodata & \nodata & \nodata & \\
        2023/08/20 & 08:07:30 & \nodata & 3.2 & \nodata & \nodata & \nodata & \nodata & \nodata & \nodata & \nodata & \nodata & \nodata & \\
        2023/08/26 & 07:49:29 & \nodata & 1.8 & \nodata & \nodata & \nodata & \nodata & \nodata & \nodata & \nodata & \nodata & \nodata & \\
        2023/08/27 & 08:25:22 & \nodata & 2.8 & \nodata & \nodata & \nodata & \nodata & \nodata & \nodata & \nodata & \nodata & \nodata & \\
        2023/08/28 & 08:25:03 & 36.5 & 1.9 & -4.23 & 94.7 & 83.6 & 2.13 & 0.92 & 0.17 & 14.7 & 137.2 & 2.93 & Possible SDA, split spectra \\
        2023/09/11 & 03:58:30 & \nodata & 1.9 & \nodata & \nodata & \nodata & \nodata & \nodata & \nodata & \nodata & \nodata & \nodata & Fe-rich \\
        2023/09/14 & 01:45:19 & 20.5 & 2.3 & -4.26 & 94.5 & 88.2 & 2.45 & 0.66 & 0.83 & 15.9 & 236.9 & 3.11 & Fe-rich \\
        2023/09/14 & 02:46:30 & 17.8 & 1.2 & -3.61 & 84.9 & 78.9 & 2.25 & 0.63 & 0.83 & 2.0 & 55.7 & 3.34 & Fe-rich \\
        2023/09/14 & 03:00:09 & \nodata & 1.3 & \nodata & \nodata & \nodata & \nodata & \nodata & \nodata & \nodata & \nodata & \nodata & Fe-rich \\
        2023/09/14 & 04:23:14 & \nodata & 2.6 & \nodata & \nodata & \nodata & \nodata & \nodata & \nodata & \nodata & \nodata & \nodata & Split spectra \\
        2023/09/16 & 01:18:35 & \nodata & 2.8 & \nodata & \nodata & \nodata & \nodata & \nodata & \nodata & \nodata & \nodata & \nodata & Fast \\
        2023/09/16 & 02:43:23 & 22.5 & 1.9 & -4.27 & 87.0 & 79.8 & 2.22 & 0.67 & 0.73 & 18.4 & 249.9 & 3.29 & Fe-rich \\
        2023/09/16 & 04:32:31 & 26.0 & 2.3 & -4.43 & 88.1 & 81.9 & 1.59 & 0.73 & 0.43 & 4.6 & 290.2 & 4.02 & \\
        2023/09/19 & 03:00:25 & 10.6 & 1.9 & -3.55 & 74.7 & 71.8 & 1.00 & 0.02 & 0.98 & 0.0 & 285.0 & 6.08 & \\
        \textbf{2023/09/19} & \textbf{03:41:59} & 21.5 & 2.0 & -4.17 & 88.6 & 83.9 & 3.46 & 0.78 & 0.76 & 3.4 & 64.2 & 2.52 & Transverse fragmentation \\
        2023/09/19 & 05:04:54 & \nodata & 1.0 & \nodata & \nodata & \nodata & \nodata & \nodata & \nodata & \nodata & \nodata & \nodata & \\
        2023/09/20 & 02:49:27 & \nodata & 1.0 & \nodata & \nodata & \nodata & \nodata & \nodata & \nodata & \nodata & \nodata & \nodata & \\
        2023/09/20 & 05:29:41 & \nodata & 2.5 & \nodata & \nodata & \nodata & \nodata & \nodata & \nodata & \nodata & \nodata & \nodata & Fe-rich \\
        2023/09/22 & 02:02:29 & 14.4 & 1.3 & -3.25 & 82.5 & 75.9 & 1.50 & 0.40 & 0.09 & 4.0 & 50.6 & 4.4 & Fe-rich \\
        2023/09/22 & 03:24:11 & 17.5 & 1.9 & -3.62 & 89.0 & 74.0 & 2.66 & 0.64 & 0.96 & 15.6 & 209.2 & 3.02 & Fe-rich \\
        2023/09/23 & 03:14:46 & \nodata & 1.6 & \nodata & \nodata & \nodata & \nodata & \nodata & \nodata & \nodata & \nodata & \nodata & Fe-rich \\
        2023/09/23 & 03:20:01 & \nodata & 2.9 & \nodata & \nodata & \nodata & \nodata & \nodata & \nodata & \nodata & \nodata & \nodata & \\
        2023/09/24 & 04:02:01 & 20.0 & 1.5 & -3.85 & 91.6 & 82.9 & 2.82 & 0.72 & 0.79 & 1.7 & 61.8 & 2.86 & \\
        2023/10/15 & 02:13:56 & 33.0 & 1.2 & -4.33 & 98.1 & 91.5 & 1.72 & 0.84 & 0.28 & 22.7 & 305.4 & 3.61 & \\
        2023/10/22 & 06:58:01 & \nodata & 3.2 & \nodata & \nodata & \nodata & \nodata & \nodata & \nodata & \nodata & \nodata & \nodata & \\
        2023/11/07 & 07:48:37 & \nodata & 2.5 & \nodata & \nodata & \nodata & \nodata & \nodata & \nodata & \nodata & \nodata & \nodata & \\
        2023/11/08 & 03:16:20 & \nodata & 1.9 & \nodata & \nodata & \nodata & \nodata & \nodata & \nodata & \nodata & \nodata & \nodata & \\
        2023/11/10 & 00:04:13 & 29.4 & 1.1 & -3.79 & 89.6 & 75.7 & 2.49 & 0.6 & 1.00 & 45.6 & 187.3 & 2.86 & \\
        2023/11/10 & 01:52:54 & \nodata & 2.8 & \nodata & \nodata & \nodata & \nodata & \nodata & \nodata & \nodata & \nodata & \nodata & \\
        2023/11/10 & 06:56:33 & \nodata & 2.4 & \nodata & \nodata & \nodata & \nodata & \nodata & \nodata & \nodata & \nodata & \nodata & \\
        2023/11/10 & 10:00:34 & \nodata & 3.0 & \nodata & \nodata & \nodata & \nodata & \nodata & \nodata & \nodata & \nodata & \nodata & \\
        2023/11/12 & 00:43:27 & \nodata & 2.2 & \nodata & \nodata & \nodata & \nodata & \nodata & \nodata & \nodata & \nodata & \nodata & \\
        2023/11/12 & 07:26:53 & 18.1 & 0.1 & -3.39 & 81.2 & 75.7 & 2.58 & 0.68 & 0.83 & 4.2 & 233.5 & 3.05 \\
        2023/11/15 & 09:39:35 & \nodata & 2.3 & \nodata & \nodata & \nodata & \nodata & \nodata & \nodata & \nodata & \nodata & \nodata & \\
        2023/11/16 & 01:19:28 & \nodata & 2.4 & \nodata & \nodata & \nodata & \nodata & \nodata & \nodata & \nodata & \nodata & \nodata & \\
        2023/11/16 & 03:40:54 & \nodata & 2.5 & \nodata & \nodata & \nodata & \nodata & \nodata & \nodata & \nodata & \nodata & \nodata & \\
        2023/11/17 & 23:52:33 & 21.0 & 1.1 & -3.72 & 88.6 & 85.5 & 2.53 & 0.69 & 0.78 & 12.0 & 61.5 & 3.04 \\
        2023/11/18 & 03:36:19 &  \nodata & 1.6 & \nodata & \nodata & \nodata & \nodata & \nodata & \nodata & \nodata & \nodata & \nodata & \\
        2023/11/18 & 06:20:38 & 17.1 & 1.0 & -4.18 & 81.4 & 78.1 & 1.08 & 0.43 & 0.62 & 3.0 & 104.5 & 5.64 & \\
        2023/11/19 & 09:02:35 &  \nodata & 2.5 & \nodata & \nodata & \nodata & \nodata & \nodata & \nodata & \nodata & \nodata & \nodata & \\
        2023/11/20 & 06:23:27 & \nodata & 1.9 & \nodata & \nodata & \nodata & \nodata & \nodata & \nodata & \nodata & \nodata & \nodata & \\
        \textbf{2023/11/20} & \textbf{09:43:04} & \nodata & 2.7 & \nodata & \nodata & \nodata & \nodata & \nodata & \nodata & \nodata & \nodata & \nodata & Transverse fragmentation \\ 
        2023/12/13 & 07:59:27 & \nodata & 2.5 & \nodata & \nodata & \nodata & \nodata & \nodata & \nodata & \nodata & \nodata & \nodata & \\
        2023/12/13 & 09:07:59 & 21.6 & 1.8 & -4.00 & 88.9 & 84.5 & 1.65 & 0.57 & 0.71 & 19.5 & 76.7 & 4.02 & \\
        2023/12/14 & 04:46:55 & \nodata & 2.0 & \nodata & \nodata & \nodata & \nodata & \nodata & \nodata & \nodata & \nodata & \nodata & \\
        \textbf{2023/12/15} & \textbf{10:53:17} & 26.4 & 2.1 & -4.00 & 89.1 & 77.5 & 1.96 & 0.76 & 0.47 & 7.6 & 101.1 & 3.45 & Distinct split spectra \\
        \textbf{2023/12/20} & \textbf{04:40:34} & 24.4 & 1.5 & -3.83 & 87.7 & 75.7 & 0.73 & 0.71 & 0.21 & 18.7 & 333.1 & 7.63 & Split spectra \\
        2024/02/04 & 00:50:43 & \nodata & 2.3 & \nodata & \nodata & \nodata & \nodata & \nodata & \nodata & \nodata & \nodata & \nodata & Distinct split spectra \\
        2024/02/04 & 03:02:56 & \nodata & 2.7 & \nodata & \nodata & \nodata & \nodata & \nodata & \nodata & \nodata & \nodata & \nodata & Fe-rich \\
        2024/02/06 & 02:26:06 & 14.5 & 1.2 & -3.86 & 81.7 & 77.9 & 1.14 & 0.295 & 0.81 & 3.8 & 258.3 & 5.44 & \\
        2024/02/06 & 05:21:30 & \nodata & 1.8 & \nodata & \nodata & \nodata & \nodata & \nodata & \nodata & \nodata & \nodata & \nodata & \\
        2024/02/07 & 10:37:21 & \nodata & 1.0 & \nodata & \nodata & \nodata & \nodata & \nodata & \nodata & \nodata & \nodata & \nodata & \\
\enddata
\end{deluxetable}
\end{longrotatetable}

\section{Discussion}

Based on a year of data since initial commissioning, preliminary analysis shows that CAMO-S as constructed meets its design goals. It has been shown to be effective at automatically collecting a large number of useable spectra even in short time intervals, as shown by the 2023 Perseids presented here. All the spectra are suitable for modelling to determine element ratios, a key design goal. It is hoped that more detailed modelling may also allow for absolute element abundances (albeit with many model assumptions). As part of this modelling effort data are already being gathered to correlate CAMO-S meteors with multi-frequency measurements of common radar echoes by the Canadian Meteor Orbit Radar (CMOR). This would allow an independent estimate of electron line densities, following the procedure described in \citet{Stober2021}.

While fragments with distinct compositions have not yet been recorded by CAMO-S, the current observations show that this is feasible. Although the spectral system is able to observe the Calcium H and K lines, which may be present if the energy input is high enough to ablate a CAI, a better indicator of the evaporation of refractory Ca-rich material may be the neutral Ca line at 423 nm. Neither of the two meteors with Ca in their spectra showed distinct fragments, though the H and K lines of Ca(II) appear later in the meteoroid ablation in both cases; this could be due to the energy needed to ablate Ca-containing minerals being present, or it could be a sign of the development of a shock wave. The presence of the neutral Ca line at 423 nm was also observed in this work, and will be the focus of a later study.

It has also been demonstrated that CAMO-S is capable of imaging spectra of separate meteoroid fragments; in the case of the iron meteors recorded here, none of the distinguishable spectra differed from fragment to fragment, suggesting that the meteoroid was chemically homogeneous or split into smaller pieces with the same overall composition. It is particularly interesting that most iron meteors indentified in the sample showed a distinctive split into several fragments, as opposed to the continuous wake of small fragments normally seen with CAMO-O. For the iron-rich meteors (those containing some Na in addition to Fe), gross fragmentation was less likely with continuous fragmentation being common. Most of the iron and iron-rich meteors for which trajectories could be computed have speeds lower than 30 km s$^{-1}$, and asteroidal or nearly asteroidal Tisserand parameters ($T_J\geq 3$), consistent with the results reported by \citet{Mills2021}. 

\section{Conclusions}

A novel mirror tracking meteor spectral recording system has been designed, built and is now in regular operation. The CAMO-S system was designed to characterize the chemical composition of faint meteors, investigate spectral differences between the head and wake of meteors, and search for inclusions with different properties from the bulk composition of the meteoroid. The preliminary observations gathered since the system became fully operational in mid-2023 shows that all the main design goals have been met. Improvements in spatial resolution and blue sensitivity are planned in future. 

It has been shown here that meteoroids with strong iron lines in their spectra fragment into distinct fragments, rather than undergoing continuous fragmentation. This may be due to splitting of liquid droplets as the iron melts. 

More detailed analysis of individual spectra will be presented in a forthcoming paper.

\section{Acknowledgements}
This work was supported in part by the NASA Meteoroid Environment Office under cooperative agreements 80NSSC18M0046, 80NSSC21M0073 and 80NSSC24M0060. MCB and PGB also acknowledge funding support from the Natural Sciences and Engineering Research Council of Canada and the Canada Research Chairs program. M. Scott and T. Mills helped developed earlier versions of MESS, and C. Baerts helped parse the events.

\appendix

\section{Spectral sensitivity}
\label{app:SpectralSensitivity}

\begin{figure}[h]
    \centering
    \includegraphics[width=\textwidth]{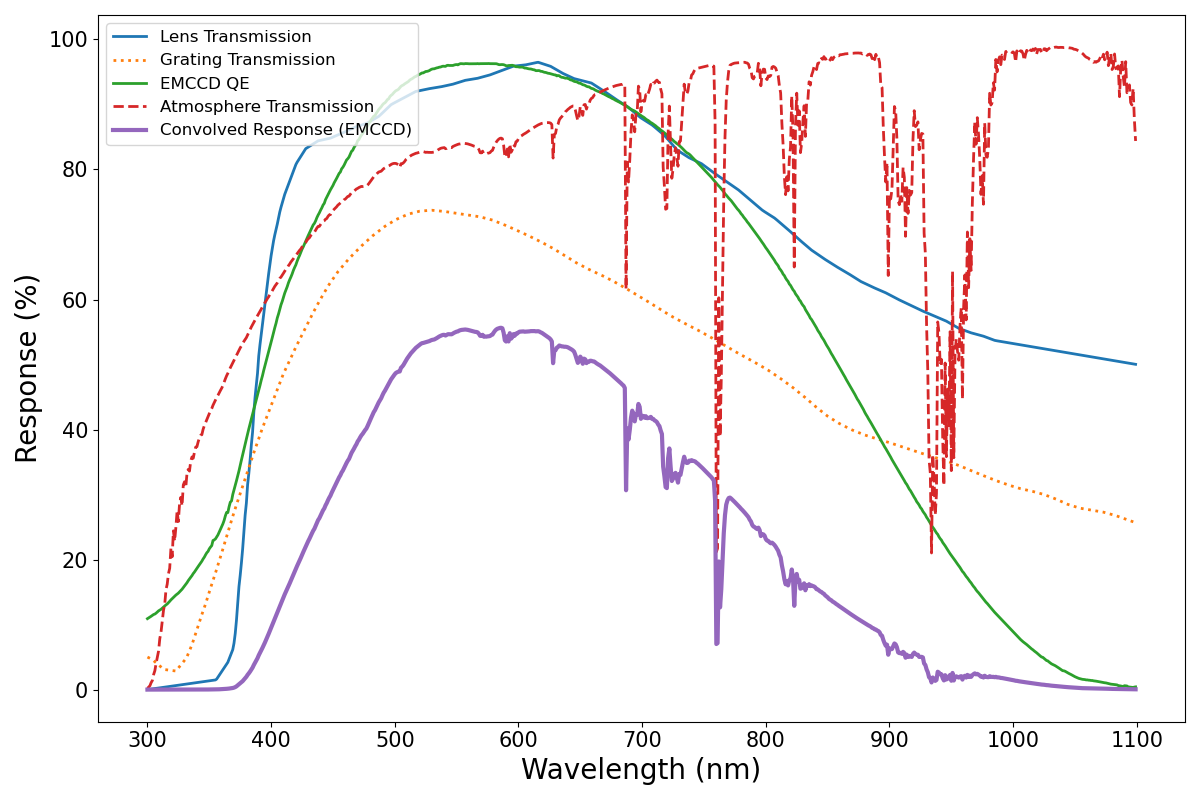}
    \caption{Transmission and sensitivities of the different elements of the CAMO-S optical system along with the theoretical convolved response for the EMCCD. Also shown for reference is the sensivitiy for a Generation 3 extended blue intensifer. The lens transmission was taken from \cite{Nagasawa2016}. The grating transmission is taken directly from the specification sheet from Edmund Optics for the 150 lpmm blazed transmission grating \footnote{\url{https://www.edmundoptics.ca/document/download/491990}}, while theintensifier quantum efficiency for the Intensified Extended Blue QImaging Retiga is from the manufacturer. The EMCCD quantum efficiency is also extracted from the manufacturer \footnote{www.nuvucameras.com/products/hnu-512/}. Finally, the atmospheric response as a function of wavelength is taken from (GIVE REFERENCE).}
    \label{fig:EstimatedResponsivityEMCCD}
\end{figure}

As with any optical system, the responsivity of CAMO-S is dependent on a number of different factors. These include the spectral response of the imaging camera, the transmission of the optical elements (lens, grating), the reflectivity of the mirrors, and the transmission of the atmosphere. These need to be quantified so that a raw (measured) spectrum can be transformed into a properly (intensity) calibrated spectrum.

The spectral response of the Nuvu EMCCD was digitized from the product literature and is specified to be sensitive from $\leq$350 nm to about 950 nm.

The transmission of the optical elements of the camera is assumed to be dominated by the transmission of the 180 mm lens. Here we use the curve from \citet{Nagasawa2016} for a similar Nikon lens as an estimate of the throughput as we have not measured this in the lab. The absolute transmission of the grating is taken from the manufacturer's specification sheet \footnote{\url{https://www.gratinglab.com/Products/Product_Tables/Efficiency/Efficiency.aspx?catalog=54-*-760R}}.

For atmospheric transmission, we assume that the top of the atmosphere coincides with the top of the mesosphere at 85 km altitude. As this is roughly the height at which we expect to see most of our meteors, we can use the TAPAS service \citep{Bertaux2014} with the US 1976 Standard atmosphere model for calculating transmission through the atmosphere at the Elginfield site for a zenith angle of 45\textdegree\ (Figure \ref{fig:EstimatedResponsivityEMCCD}). 

An approximation of the total CAMO-S responsivity (Figure \ref{fig:EstimatedResponsivityEMCCD}) is made by convolving these individual components with one another. This is then used as a starting point for the generation of the final responsivity curve. This baseline responsivity shows a low response of approximately 10$\%$ for the region around the high-temperature calcium H and K lines (393-397 nm) an increased response of about 50$\%$ for the low-temperature iron and magnesium multiplets between 500 and 600 nm and the atmospheric oxygen lines between 530 and 700 nm. At wavelengths longer than 700 nm, the response is largely impacted by atmospheric absorption, which can attenuate the observed strength of the meteor lines. As a result, we expect to see highly attenuated spectral features below about 400 nm and beyond 900 nm, with the most pronounced lines being visible between 500 and 700 nm.

In practice, finding the true end-to-end responsivity of the system is more involved than the simple convolutional assumptions made above. We start by pointing the system toward a calibration star and imaging the spectrum for 20 seconds at a number of different mirror offsets in both the x and y directions. This allows us to not only reconstruct the stellar spectrum over a large wavelength range (about 375 nm to 1 $\mu$m) but to also examine the influence of field position on the spectral scale and responsivity. The individual frames in each 10-second video file are combined to create a mean stack before flat-field removal. Flat fields are generated by pointing to a star-free region of the sky and slowly panning around the field while collecting raw video files of approximately 30 seconds in length. The individual frames from the videos are processed to create a set of mean stacks (one for each video) which are median combined to produce the final flat-field image  used to correct the spectral data. 

After flat-field removal, the images are rotated to align the spectra along rows before shifting and stacking with other images with similar mirror offsets to further reduce noise. Once this is done, we extract the mean spectrum from each mirror offset and then align them based on their spectral features. The observed absorption features are then matched with the same features on the known spectrum to calculate the spectral scale.

As this is an involved procedure we have so far only done this with one bright star visible in our north pointing direction, namely $\delta$ UMi. We were able to identify prominent hydrogen Balmer lines (H-$\gamma$, H-$\delta$, H-$\beta$, and H-$\alpha$) absorption features and calculate a spectral scale of  0.468 nm/pixel at full sensor resolution. It follows that for a 2x2 binned image we would expect a spectral scale of 0.935 nm/pixel with the caveat that any distortion in the lens could change the scale when moving from center to edge of the field.

\bibliography{camo-spectral}
\bibliographystyle{aasjournal}



\end{document}